# Why comparing survival curves between two subgroups may be misleading

Damjan Krstajic

Research Centre for Cheminformatics, Jasenova 7, 11030 Beograd, Serbia

Email address: damjan.krstajic@rcc.org.rs

## Abstract

We analyse an issue when comparing survival curves between two subgroups. We show that there is a direct relationship between estimates of subgroups' survival at a time point and positive and negative predictive values in the binary classification settings. Our findings present a case where current methods of comparing survival curves between subgroups may be misleading. We think that this ought to be taken into account during the validation of prognostic diagnostic tests that predict two prognostic subgroups for a given disease or treatment, when the validation data set consists of censored data.

## Introduction

The goal of personalised medicine is to apply treatments only to individuals who will benefit from them. This presumes that we have available information about the patients as well as statistical methodology to identify a subgroup which benefits from a treatment. From the statistical point of view, the future of personalised medicine depends on how well we can separate a subgroup of patients that benefits from a specific treatment. Our starting point is a single survival population. Our goal is to find two subgroups, one with better and the other with worse survival prognoses. However, in order to achieve that goal we need to have a set of well-defined statistics that measure, as accurately as possible, the differences between the subgroups' survival prognoses.

A common procedure when validating prognostic separation is to perform the same actions as when comparing two survival populations, i.e. to estimate survivor functions for each subgroup with the Kaplan-Meier (KM) survival curve [1], to calculate the hazard ratio (HR) [2] and to compare the curves with the log-rank test [3]. For example, Royston and Altman [4] say the following on this subject: "*Kaplan-Meier survival curves for risk groups provide informal evidence of discrimination. The more widely separated are the curves, the better is the discrimination. A Kaplan-Meier graph for both datasets allows a visual comparison of discrimination between datasets. We strongly recommend producing such plots.*".

Our thesis is that comparing survival curves between two subgroups may be misleading in the same way as when evaluating a binary classifier based on its predictive values without taking into account the prevalence.

## Methods

It is very important that we clearly define the terms we use. The *survival population* is a population of individuals each having a “failure time”. Let $T$ be a non-negative random variable representing the “failure time” of an individual from the survival population. The *survivor function* $S(t) = P(\{T>t\})$ is the probability that $T$ exceeds the value t [5][6]. The *population survival rate* at time To S(To) is the proportion of the population with “failure time” after To.

We presume that we have a rule that divides subjects into two groups A and B. We are interested to find whether the observed survival of subjects in groups A and B differ. Lets assume that we expect that subjects from group A have better chances of survival than subjects from group B. Sometimes in practice the subgroups are created after dichotomising a prognostic index, a numerical output of a prognostic model [4].

We are interested in the process of *external validation*, which means assessing the group A and B separation on an independent dataset [4]. We refer to a dataset used in external validation as the *validation dataset*. In practice we usually do not know the survivor functions, but we are able to estimate them. We use the term *survival curve* $\hat{S}(t)$ when we refer to an estimate of a survivor function. The *validation survival rate* at time To $\hat{S}(To)$ is the proportion of the validation dataset with “failure time” after To.

In order to explain our point we will first suppose that there is no censored data in our validation dataset, i.e. we know the exact failure time for each subject. We will follow up with arguments that the same applies for cases with censored data.

## Estimating the survival of subgroup’s patients at a time point in the absence of censoring

To start with, we are interested in estimating the survival of subjects in subgroups A and B at time point To, i.e. $\hat{S}_A(To)$ and $\hat{S}_B(To)$. In the absence of censoring, we can estimate it for each subgroup as a ratio of the number of subjects who had an event after To over the size of the subgroup in the validation dataset. Therefore, if we categorise patients with an event before or equal to To as *positive*, and with an event after To as *negative*, then we can create a contingency table as in Table 1. In that case $\hat{S}_A(To)=a/(a+b)$ and $\hat{S}_B(To)=c/(c+d)$ and we refer to them as *naive estimates*.

| | NEGATIVE | POSITIVE |
|---|---|---|
| subgroup A | a | b |
| subgroup B | c | d |

Table 1 - A contingency table between positive and negative patients and subgroup A and B patients

In the absence of censoring, the separation of the population into two subgroups may be seen as a classification problem. We introduce the following convention. Subjects who had an event after To and who are in subgroup A would be true negatives (TN), while those who are in subgroup B would be false negatives (FN). Similarly, subjects with an event before or equal to To who are in subgroup B would be true positives (TP), while those who are in subgroup A would be false positives (FP). Using this simple diagnostic test we can calculate sensitivity, specificity, positive predictive value (PPV) and negative predictive value (NPV) [7][8]. We can estimate the sensitivity and specificity from a validation set, but without the knowledge of prevalence we are unable to estimate PPV or NPV [8]. The prevalence is the proportion of positive patients in the population, also known as the prior probability of being positive. It is a measure independent of the validation dataset. If, however, we know the prevalence, then the predictive values would be estimated with the following equations [8]:

$$PPV = \frac{sens \cdot prev}{sens \cdot prev + (1 - spec) \cdot (1 - prev)} \quad \text{(Eq1)}$$

$$NPV = \frac{spec \cdot (1 - prev)}{spec \cdot (1 - prev) + prev \cdot (1 - sens)} \quad \text{(Eq2)}$$

In our case then $\hat{S}_A(To)$ and $\hat{S}_B(To)$ would be estimates of proportions of negative samples in subgroups A and B, i.e.

$$\hat{S}_B(To) = (1 - PPV) \quad \text{(Eq3)}$$

$$\hat{S}_A(To) = NPV \quad \text{(Eq4)}$$

And here we have a problem. Their estimates depend on the prevalence. In our case prevalence is the proportion of subjects in the population who had an event before or equal to time To, i.e.

$$\text{prevalence} = (1 - S(To)) \quad \text{(Eq5)}$$

Therefore, we cannot estimate $\hat{S}_A(To)$ or $\hat{S}_B(To)$ without the knowledge of the population survival rate at To, i.e. S(To). Consequently, we cannot use the naive estimates a/(a+b) and c/(c+d) as our estimates for $\hat{S}_A(To)$ and $\hat{S}_B(To)$ without the knowledge of S(To).

In other words, if the prevalence is different from $(1 - \hat{S}(To))$, then naive estimates for $\hat{S}_A(To)$ and $\hat{S}_B(To)$ will be biased.

In survival settings the most common method of estimating survival at time To is to use Kaplan-Meier product-estimator [1]. However, in the absence of censoring, the KM curve estimates for both subgroups at To are identical to the naive estimates. We thus conclude that in the absence of censoring we cannot use the KM curve to estimate the survival of prognostic subgroups at To because it does not take into account the prevalence. Furthermore, as this holds for any time point, we also conclude that in the absence of censoring we cannot use KM curves to compare the survival of two subgroups.

## Difference in the survival of two subgroups at a time point in the absence of censoring

In the equation Eq6 we show that in the absence of censoring the difference between naive estimates of $\hat{S}_A(To)$ and $\hat{S}_B(To)$ does depend on the prevalence.

$$\hat{S}_A(To) - \hat{S}_B(To) = PPV + NPV - 1$$

$$\hat{S}_A(To) - \hat{S}_B(To) = \frac{prev \cdot (1 - prev) \cdot (spec + sens - 1)}{[prev \cdot (1 - sens) + spec \cdot (1 - prev)] \cdot [sens \cdot prev + (1 - spec) \cdot (1 - prev)]} \quad \text{(Eq6)}$$

However, it is not clear from the equation how the prevalence affects the difference and to what extent. Taking into account Eq5, i.e. that the prevalence is one minus the population survival rate at time To, in Fig. 1 we show the difference between $\hat{S}_A(To)$ and $\hat{S}_B$ (To) as the function of the population survival rate where specificity and sensitivity are constants.

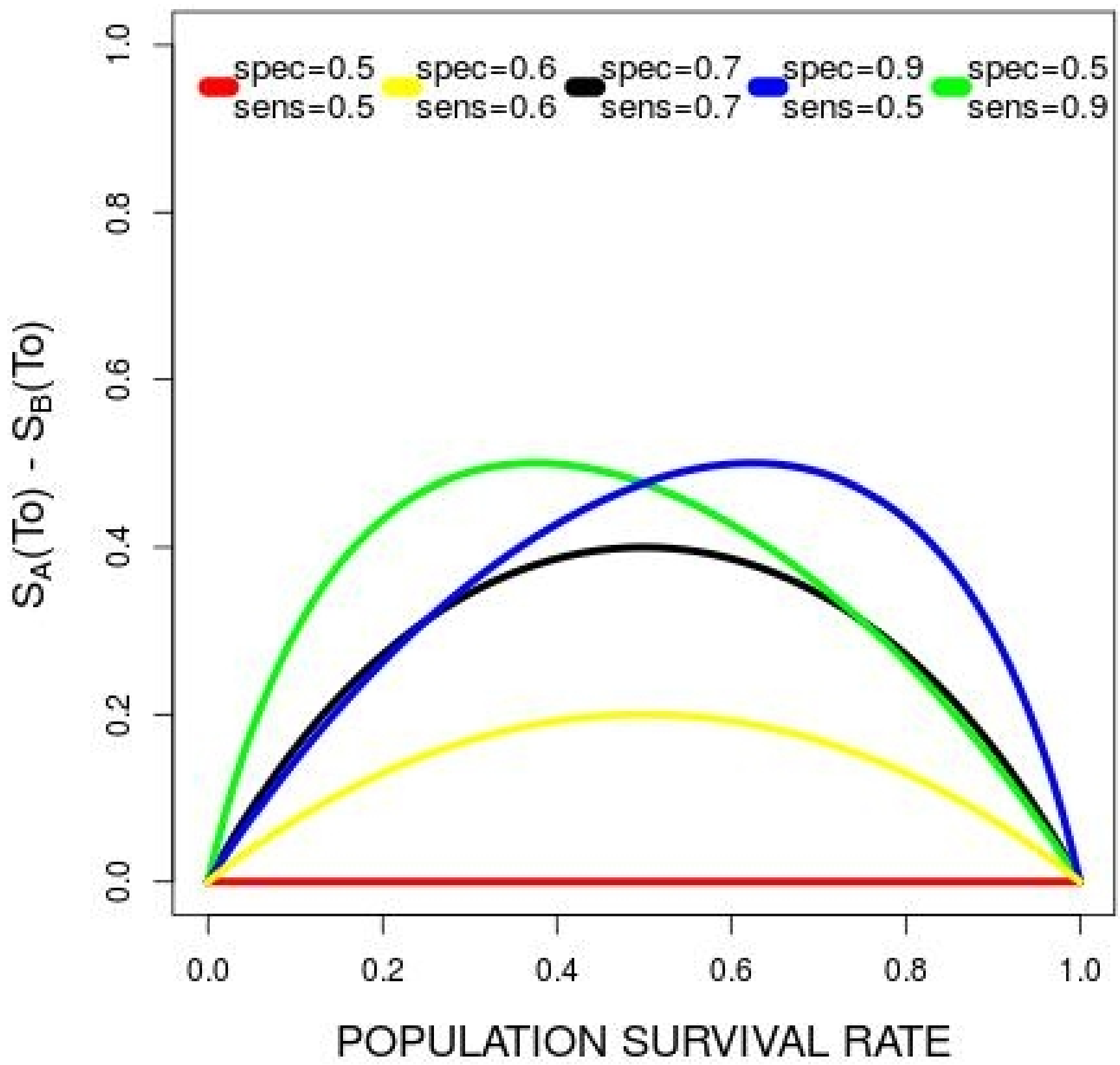

Figure 1 - Difference between $S_A$ (To) and $S_B$ (To) as the function of the population survival rate for five pairs of (sensitivity, specificity)

As it may be seen from Fig. 1, the difference between $\hat{S}_A$ (To) and $\hat{S}_B$ (To) is equal to zero when spec+sens=1. However, in other cases, the difference has a single maximum value. Furthermore, as the population survival rate approaches 0 or 1 the difference is closer to zero.

The equation Eq6 provides us with an opportunity to analyse different scenarios when the population survival rate at time To, S(To), has a different value from the validation survival rate at time To, $\hat{S}$(To). In Table 2, for each combination of (sensitivity, specificity, S(To), Ŝ(To)) we show estimated differences in subgroups' survival at time To with and without taking into account S(To). In the last column we show how biased our estimation would be if we do not take into account S(To). In our opinion, the bias in some cases is substantial.

| spec | sens | S(To) | Ŝ(To) | $S_A$(To) – $S_B$(To) | $\hat{S}_A$(To) – $\hat{S}_B$(To) | abs(($S_A$(To) – $S_B$(To)) – ($\hat{S}_A$(To) – $\hat{S}_B$(To))) |
|---|---|---|---|---|---|---|
| 0.5 | 0.9 | 0.2 | 0.4 | 0.4336 | 0.499 | 0.0654 |
| 0.5 | 0.9 | 0.5 | 0.1 | 0.4762 | 0.299 | 0.1772 |
| 0.6 | 0.6 | 0.1 | 0.3 | 0.0739 | 0.1691 | 0.0952 |
| 0.6 | 0.6 | 0.4 | 0.25 | 0.1923 | 0.1515 | 0.0408 |
| 0.7 | 0.7 | 0.3 | 0.4 | 0.3448 | 0.3865 | 0.0417 |
| 0.7 | 0.7 | 0.2 | 0.5 | 0.2716 | 0.4 | 0.1284 |
| 0.8 | 0.6 | 0.15 | 0.35 | 0.2053 | 0.3663 | 0.161 |
| 0.8 | 0.6 | 0.45 | 0.25 | 0.4064 | 0.3 | 0.1064 |
| 0.8 | 0.8 | 0.3 | 0.4 | 0.5348 | 0.5844 | 0.0496 |
| 0.8 | 0.8 | 0.2 | 0.5 | 0.4412 | 0.6 | 0.1588 |
| 0.9 | 0.5 | 0.25 | 0.45 | 0.3125 | 0.455 | 0.1425 |
| 0.9 | 0.5 | 0.55 | 0.35 | 0.4911 | 0.395 | 0.0961 |
| 0.9 | 0.9 | 0.3 | 0.4 | 0.7487 | 0.7882 | 0.0395 |
| 0.9 | 0.9 | 0.2 | 0.5 | 0.6653 | 0.8 | 0.1347 |

Table 2 - Estimated differences in subgroups' survival at time To based on the combination of (sensitivity, specificity, S(To), Ŝ(To))

## Comparing survival curves of two subgroups with censored data

We think that our conclusion holds when a validation dataset is with censored data.

We have shown that using KM curves to compare the survival of two subgroups in the absence of censoring may provide misleading and substantially biased estimates. We cannot find any argument which would suggest that comparison with censored data may prove to be less misleading or with less substantially biased estimates.

## Discussion

By analysing the difference in the survival of two subgroups at a time point, we have shown that if the validation survival rate at the time point is different from the population survival rate at the time point, we may obtain substantially biased results.

In theory, an external validation dataset ought to be representative of the population, but in practice it seldom is, or it is difficult to prove that it is. Therefore, if we want to compare survival curves of subgroups, the validation dataset ought to have the same survival rates as the population for all time points. In our opinion, this is not practical. Therefore, we question the common practice of using KM survival curves for subgroups to provide evidence of their discrimination.

In the same way that one cannot estimate PPV nor NPV without the knowledge of prevalence [8], one cannot estimate the difference in survival of subgroups at a time point without the knowledge of the population survival rate at the time point. Consequently, if we don't know the population survival rate at any time point, we cannot estimate the difference in survival of subgroups using the validation dataset alone.

## Conclusion

We have shown that the relationship between the survival of subgroups at a time point and the population survival rate at the time point is the same as between predictive values and prevalence in the classification setting. Therefore, we conclude that we may err when estimating the survival of subgroups solely by using the validation data.

## Acknowledgments

The author would like to thank his mother, Linda Louise Woodall Krstajic, for correcting English typos and language improvements in the text.